\documentclass[draftclsnofoot, onecolumn]{IEEEtran}
\IEEEoverridecommandlockouts
\usepackage{cite}
\usepackage{amsmath,amssymb,amsfonts}
\usepackage{algorithmic}
\usepackage{graphicx}
\usepackage{textcomp}
\usepackage{xcolor}
\usepackage{tabularx}
\usepackage{datatool}
\usepackage{float}

\def\BibTeX{{\rm B\kern-.05em{\sc i\kern-.025em b}\kern-.08em
    T\kern-.1667em\lower.7ex\hbox{E}\kern-.125emX}}

\begin{document}
\newcolumntype{L}[1]{>{\raggedright\arraybackslash}p{#1}}
\newcolumntype{C}[1]{>{\centering\arraybackslash}p{#1}}
\newcolumntype{R}[1]{>{\raggedleft\arraybackslash}p{#1}}

\title{\large \textbf{Mapping Trafficking Networks: A Data-Driven Approach to Disrupt Human Trafficking Post Russia-Ukraine Conflict}}

\makeatletter
\newcommand{\linebreakand}{
  \end{@IEEEauthorhalign}
  \hfill\mbox{}\par
  \mbox{}\hfill\begin{@IEEEauthorhalign}
}


\makeatother
\author{
    Murat Ozer\textsuperscript{1}, 
    Goksel Kucukkaya\textsuperscript{1}, 
    Yasin Kose\textsuperscript{2}, 
    Assel Mukasheva\textsuperscript{3}, \\
    Kazim Ciris\textsuperscript{1}, \\
    Bharath V. Penumatcha\textsuperscript{1} \\[1em]
    \textsuperscript{1}\textit{School of Information Technology, University of Cincinnati, Cincinnati, Ohio, USA} \\
    \textsuperscript{2}\textit{Cybercrime and Forensic Computing, Friedrich-Alexander-Universität, Erlangen, Germany} \\
    \textsuperscript{3}\textit{Information Systems, Kazakh-British Technical University, Almaty, Kazakhstan} \\
    \textsuperscript{1}\textit{School of Information Technology, University of Cincinnati, Cincinnati, Ohio, USA} \\ [1em]
    m.ozer@uc.edu, kucukkgl@ucmail.uc.edu, yasin.koese@fau.de, a.mukasheva@kbtu.kz, ciriskm@ucmail.uc.edu, penumaba@mail.uc.edu
}

\maketitle

\thispagestyle{plain}
\pagestyle{plain}

\begin{abstract}
This study proposes a prototype for locating important individuals and financial exchanges in networks of people trafficking that have grown during the conflict between Russia and Ukraine. It focuses on the role of digital platforms, cryptocurrencies, and the dark web in facilitating these operations. The research maps trafficking networks and identifies key players and financial flows by utilizing open-source intelligence (OSINT), social network analysis (SNA), and blockchain analysis. The results show how cryptocurrencies—Bitcoin in particular—are used for anonymous transactions and imply that upsetting central coordinators may cause wider networks to become unstable. In order to combat human trafficking, the study emphasizes the significance of real-time data sharing between international law enforcement. It also identifies future directions for the development of improved monitoring tools and cooperative platforms. 
\end{abstract}
 
\section{Introduction}
The Russia-Ukraine conflict, which started in early 2022, has made human trafficking, which is already a global problem, worse. Millions of people have been forcibly relocated, mainly women, children, and the elderly, leaving vulnerable groups open to abuse by human traffickers. Traffickers take advantage of displaced people who have no family or means of support because of the unrest, poor governance, and weak law enforcement in conflict areas. These risks have been heightened by the war, as widespread migration throughout Europe and Ukraine has left a vulnerable refugee population. The confusion created by traffickers is used to coerce victims into labor or sexual exploitation, underscoring the critical need for early, technologically advanced interventions.

Trafficking activities are now transnational in nature. The emergence of illicit online networks, especially those functioning on the dark web, and digital technologies has allowed traffickers to reach a wider audience worldwide, facilitating the recruitment and exploitation of victims across national boundaries. Traffickers use online platforms to facilitate the buying and selling of human lives, often utilizing cryptocurrency transactions to maintain anonymity and evade law enforcement \cite{gao2022}.  

Millions of people are seeking asylum in nearby nations like Germany and Poland as a result of the biggest refugee crisis in Europe since World War II, which was sparked by the conflict between Russia and Ukraine. Even though a large number of people have received aid, the sheer volume of displaced people has put a strain on local resources and left refugees open to abuse. Human traffickers pose as aid workers or make false promises of safety, according to reports from organizations like the UN and Europol, which indicate a sharp increase in cases of human trafficking \cite{unodc2023report}. Trafficking networks have grown both online and offline, operating through encrypted channels, dark web platforms, and cryptocurrency. Because of this sophistication, it is challenging for law enforcement to step in without using thorough, data-driven plans. \cite{unodc2022}.

In light of this, our research uses an interdisciplinary methodology to disentangle the intricate web of trafficking operations by combining social network mapping, blockchain analysis, and open-source intelligence. By doing this, we hope to uncover possible weaknesses that could be used to interfere with these networks' operations and throw light on the complex mechanisms that keep them running. This study's implications go well beyond the immediate context of the conflict between Russia and Ukraine. Lessons from this research can help develop more effective strategies to combat human trafficking in other war-torn regions as conflicts continue to break out around the world. Furthermore, our results highlight how important it is for nations to work together and use cutting-edge technology to combat this widespread violation of human rights.  

\subsection{\textbf{Research Objectives and Significance}}

This study's main goal is to disentangle the complex web of human trafficking activities that have grown in response to the conflict between Russia and Ukraine. The objective of this study is to provide light on the ways in which traffickers are exploiting vulnerable populations through the evolution of their tactics by examining both the digital and physical aspects of trafficking. In particular, we aim to investigate (1) how trafficking in conflict areas is facilitated by the dark web, (2) the part cryptocurrencies play in helping traffickers stay anonymous, and(3) the patterns and tendencies in exploitation, transportation, and recruitment since the commencement of the conflict between Russia and Ukraine.  

The study will map these networks and investigate how contemporary technologies, like open-source intelligence, blockchain analysis, and real-time data sharing, can be used to thwart trafficking operations and aid investigations. The findings will inform both local and international law enforcement agencies, helping them to develop targeted interventions that address the root causes and operational methods of human trafficking in the digital age.  

\begin{figure}
    \centering
    \includegraphics[width=1\columnwidth]{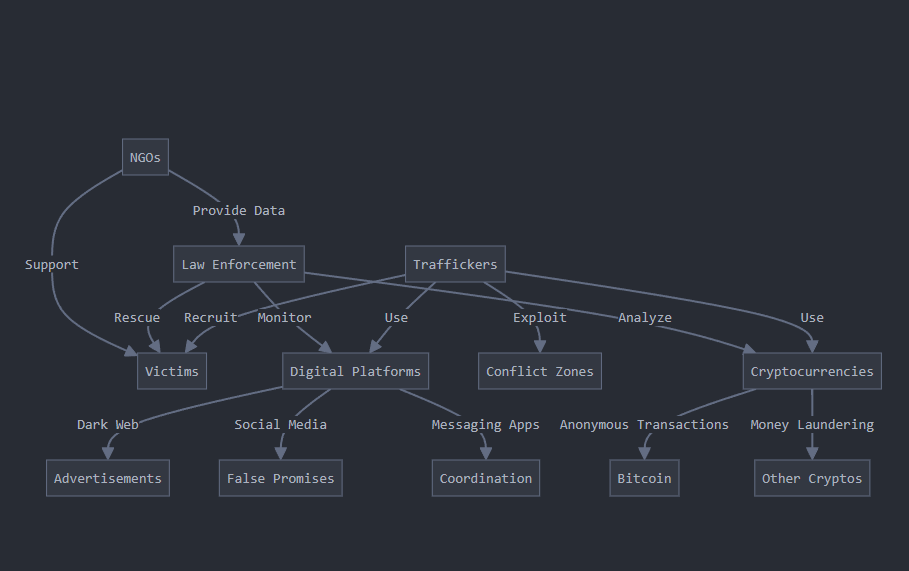}
    \caption{Network Diagram  }
    \label{fig1}
\end{figure}

The intricate and diverse character of human trafficking operations is exemplified by the network diagram, especially when considering the Russia-Ukraine conflict. Traffickers, positioned at the center, take advantage of the weaknesses brought about by conflict to recruit and control victims through a combination of traditional and digital methods. Digital platforms like social media, messaging apps, and dark web advertisements make recruitment, coordination, and increased anonymity easier. The use of cryptocurrencies—such as Bitcoin—allows for anonymous financial transactions, which increases complexity. The diagram highlights the need for sophisticated, technologically driven responses and emphasizes the crucial role that NGOs and law enforcement play in fighting these networks.

This interpretation highlights the intricacy of contemporary trafficking operations and the numerous stakeholders involved in both committing and preventing these crimes by offering a summary of the major components and relationships shown in the network diagram.  

\subsection{Theoretical Framework }

This study's theoretical foundation is based on the systemic nature of human trafficking, in which traffickers function within dispersed, frequently covert networks. Similar to the drug trade, trafficking grows in places with little control and enforcement, especially during times of chaos like war. This study examines the relationships between traffickers, victims, and intermediaries using social network analysis (SNA), which is informed by criminological theories of organized crime and digital illicit networks \cite{papachristos2012social}. In addition, the research will utilize a multidisciplinary methodology, combining perspectives from the fields of criminal justice, information technology, and international relations to investigate the technological and human aspects of contemporary human trafficking.  

\section{Relevant Studies}
\textbf{Human Trafficking and Conflict Zones }

Conflict areas have long been linked to human trafficking because of the favorable conditions for exploitation they provide due to deficiencies in social infrastructure, law enforcement, and governance. Studies reveal that vulnerable groups including refugees, internally displaced people, and unaccompanied minors are more likely to be trafficked for forced labor, sexual exploitation, and other forms of abuse during times of war and civil unrest \cite{shelley2010human}. Since chaos and disorder make it easier for traffickers to operate undetected, conflict-induced displacement gives them access to new victims. Shelley (2010) carried out one of the most thorough studies on human trafficking during conflict, looking at trafficking patterns in sub-Saharan Africa, Afghanistan, and the Balkans. Shelley's work demonstrates how people trafficking exploits those who are displaced and impoverished by war, following the patterns of international conflicts \cite{shelley2010human}. This study serves as a basis for comprehending the systemic nature of human trafficking in conflict environments and offers a context for analyzing current instances such as the conflict between Russia and Ukraine.   

Heidbrink (2014) also investigated the connection between human trafficking and conflict, concentrating on child migrants in conflict areas \cite{heidbrink2014migrant}. Heidbrink's research revealed how children who have been uprooted by conflict are regularly the focus of human trafficking, frequently coerced into labor, or seduced by armed factions. This study is especially pertinent in light of the thousands of displaced children in Ukraine, who are now more susceptible to human trafficking.   

\textbf{Human Trafficking in Eastern Europe }

Since the fall of the Soviet Union, human trafficking has been a persistent problem in Eastern Europe, especially in Ukraine. Following its independence in 1991, Ukraine developed into a significant hub for human trafficking, serving as both a source and a destination. Economic volatility, high jobless rates, and pervasive corruption fostered a climate that was conducive to human trafficking \cite{vogt2019trafficking}. Prior to the current crisis with Russia, Ukraine was known to be a major hub for forced labor and sexual exploitation as well as human trafficking.  

Vogt's (2019) study on human trafficking in Ukraine revealed that traffickers moved victims throughout Eastern Europe and into Western European nations by taking advantage of the nation's open borders and dishonest legal systems. This study is important because it shows how trafficking networks were already well-established prior to the start of hostilities, which is important for understanding the pre-war context of trafficking in Ukraine.  

Similar to this, Brunovskis and Surtees (2008) investigated the ways in which traffickers recruit, transport, and take advantage of victims in Eastern Europe, including Ukraine \cite{brunovskis2008agency}. They discovered that traffickers frequently entice victims with false promises of marriage or work overseas, only to ensnare them in prostitution or forced labor. This is particularly important given the ongoing conflict in Ukraine, where a large number of displaced women and children have fled, raising the possibility that they will fall victim to human traffickers posing as help seekers.  

\textbf{Human Trafficking and Digital Platforms }

The use of digital platforms, especially the dark web, to facilitate recruitment, exploitation, and transactions has been one of the major developments in human trafficking over the past 20 years \cite{latonero2012rise}. In order to conduct their operations and avoid detection by law enforcement, traffickers have embraced social media platforms, encrypted communication channels, and online technologies.  

According to Latonero's groundbreaking research on technology and human trafficking, traffickers use the internet to lure victims in with false marriage proposals, job offers, and other fraudulent schemes. His research focuses on the ways in which digital platforms have altered the trafficking landscape, facilitating traffickers' ability to interact, conduct business, and broaden their networks internationally.  

Christin (2013) also looked at how the dark web helps with illegal activity, such as human trafficking \cite{christin2013traveling}. According to Christin's analysis, traffickers can advertise their services and exchange payments using cryptocurrencies in a safe and anonymous environment on dark web marketplaces like Silk Road. Understanding how traffickers in conflict areas, like the Ukraine, use the dark web to stay anonymous and access a worldwide market for human exploitation is made possible by the findings of this study.  

Furthermore, McGuire and Dowling (2013) looked into how cryptocurrencies fit into the human trafficking industry. According to their research, traffickers are using cryptocurrencies—Bitcoin in particular—more frequently to hide their financial activities and avoid being discovered. This is especially important in conflict areas where traffickers need anonymous, safe ways to make payments in order to elude detection by authorities \cite{mcguire2013cybercrime}.  

\textbf{The Role of Cryptocurrency in Human Trafficking }

Because cryptocurrencies can allow human traffickers to operate in secret, there has been an increasing amount of attention paid to this practice in recent years. Because cryptocurrency transactions are decentralized and frequently anonymous, especially on blockchain platforms, they give traffickers the ability to transfer money without the supervision of conventional banking institutions \cite{meiklejohn2013fistful}. A seminal study on the use of Bitcoin for illicit transactions was conducted by Meiklejohn et al. (2013), who demonstrated how transactions using cryptocurrencies can be tracked down but are still challenging to assign to particular people. Given that traffickers may use cryptocurrencies to finance their operations and transfer profits across borders, this study is especially pertinent to efforts to combat human trafficking in the context of the Russia-Ukraine war.

Foley et al. (2019) have conducted more recent research on the use of blockchain analysis by law enforcement to track cryptocurrency transactions and locate trafficking networks \cite{foley2019sex}. According to Foley's research, it may be possible to use sophisticated data analytics to identify trends in bitcoin transactions and connect those patterns to human trafficking activities. The methodology for the proposed research, which aims to use comparable techniques to track down human trafficking operations in Eastern Europe, will be informed by this study.  

\textbf{Human Trafficking and the Russia-Ukraine War }

Human trafficking cases have significantly increased since the start of the conflict between Russia and Ukraine, especially when it comes to displaced women and children (United Nations Office on Drugs and Crime \cite{unodc2023report}. The conflict has caused the greatest refugee crisis to hit Europe since World War II, uprooting millions of Ukrainians and increasing their vulnerability to human trafficking. According to a UNODC report from 2023, trafficking networks have grown significantly since the conflict began. Traffickers take advantage of the chaos caused by war to entice victims with promises of aid or job opportunities. The agency's findings demonstrate the critical need for focused interventions that address the changing tactics of human traffickers, who are increasingly reliant on cryptocurrencies and digital platforms to support their operations.

Similar to this, Europol (2023) released a report describing how organized crime networks have adjusted to the conflict in Ukraine by utilizing digital tools to facilitate cross-border people trafficking \cite{europol2023socta}. The results of Europol's investigation highlight the value of technological innovation and international collaboration in identifying and dismantling trafficking networks that take advantage of war refugees.

The corpus of research on conflict zones, human trafficking, and the role of technology in enabling illegal activity offers a strong basis for comprehending how the Russia-Ukraine war has affected human trafficking. The increased efficiency and anonymity that traffickers have been able to operate with thanks to the integration of digital tools, cryptocurrencies, and dark web platforms has rendered traditional law enforcement methods less effective. 

\section{Methodology}

This review of the literature highlights the necessity of a thorough, technologically advanced strategy to stop human trafficking in the digital era. The current research, which draws from these studies, aims to advance knowledge by investigating how blockchain technology, dark web investigations, and modern data analytics can be used to track down and disrupt the trafficking networks that have emerged from the Russia-Ukraine war. By combining social network analysis (SNA), blockchain tracing, and open-source intelligence (OSINT), this multidisciplinary study aims to dissect the networks of human trafficking that have grown or surfaced as a result of the Russia-Ukraine war. The methodology focuses on how digital platforms and cryptocurrencies help traffickers avoid detection, addressing both the digital and physical aspects of contemporary trafficking operations.

The research is divided into three primary sections: (1) gathering data; (2) analyzing data with sophisticated computational methods; and (3) validating results by working with law enforcement. Understanding the intricate nature of the networks of people trafficking that have grown out of the Russia-Ukraine conflict requires an understanding of each component.  

\subsection{Research Design}

The study employs a mixed-methods approach, incorporating both qualitative and quantitative data to offer a thorough understanding of networks involved in human trafficking. While the quantitative data will come from publicly accessible blockchain transactions, dark web platforms, and law enforcement databases, the qualitative data will be obtained through interviews with law enforcement officials, non-governmental organizations, and experts on human trafficking.   

The principal aims are (1) to pinpoint the main trafficking networks that have surfaced after the war, (2) to use cryptocurrency analysis to track down financial transactions that facilitate human trafficking, (3) to map the relationships between traffickers, victims, and intermediaries using social network analysis, and (4)to augment the examination of physical and digital trafficking networks with OSINT.  

\subsection{Data Collection}

\subsubsection{Blockchain Data for Cryptocurrency Transactions}

The study will use blockchain analysis to track transactions connected to trafficking operations because of the increase in the use of cryptocurrencies to aid in human trafficking. Traffickers now frequently utilize cryptocurrencies, especially Bitcoin, as a means of carrying out anonymous financial transactions \cite{meiklejohn2013fistful}. Blockchain technology maintains an unchangeable public ledger of every transaction, which can be examined to identify trends suggestive of illegal activity, even though cryptocurrencies provide anonymity. Using a local node, download and analyze the Bitcoin public ledger is the first step in the data collection process. We will make use of publicly accessible resources like real-time blockchain data access APIs and Blockchain Explorer (https://www.blockchain.com/explorer). The study will map connected wallets using graph theory techniques in order to identify clusters of suspicious transactions linked to human trafficking.

\subsubsection{Dark Web and Open-Source Data Collection}

To complement the blockchain analysis, we will conduct extensive data mining on the dark web, where traffickers often advertise services and communicate with buyers \cite{christin2013traveling}. Dark web markets, like Silk Road and its more recent incarnations (Dream Market, for example), are well-known centers for illegal activities, including the trafficking of human beings. We will use Tor-based crawlers, like TorBot (https://github.com/DedSecInside/TorBot), an open-source program made to harvest content from the dark web associated with illegal activity. This tool will be used to keep an eye on message boards, marketplaces, and forums where there may be signs of human trafficking. We will use OSINT techniques to gather more intelligence from social media networks frequently used by traffickers, such as Telegram, Twitter, and others, in order to increase the effectiveness of this strategy. In order to create a database of possible leads on trafficking operations, this data will be combined with information from other sources and blockchain technology.

\subsubsection{Law Enforcement and NGO Data}

In order to obtain case files, arrest records, and victim databases, we will work with law enforcement organizations and non-governmental organizations that are active in Eastern Europe. These groups own important databases with details on well-known trafficking rings, prominent suspects, and victim exploitation trends. These organizations will help us add practical insights to the blockchain and OSINT data so that we can more effectively identify human traffickers operating in both the physical and digital spheres \cite{latonero2012rise}.

\subsection{Data Analysis}

\subsubsection{Blockchain Tracing and Cryptocurrency Analysis}

The tracking of financial flows linked to human trafficking will be the main objective of the blockchain data analysis. For the purpose of tracking illegal cryptocurrency transactions, we will make use of cutting-edge blockchain tracing tools like Chainalysis and Elliptic, which are frequently employed by law enforcement. By grouping transactions and connecting them to well-known dark web markets, illicit services, and questionable activities, these tools enable the de-anonymization of blockchain addresses \cite{foley2019sex}. We will map the connections between various cryptocurrency wallets using graph analysis techniques in order to find nodes with high transactional activity, unusual transaction patterns, and large transaction flows that may point to the involvement of trafficking operations. In order to create a complete picture of the financial flows within the trafficking network, additional analysis utilizing both historical and real-time transaction data will be carried out once a suspicious transaction or address is discovered.

 \subsubsection{Social Network Analysis (SNA)}

Social network analysis (SNA) is a crucial component of this study, as it allows us to map the relationships between traffickers, victims, intermediaries, and other actors involved in trafficking networks. We will apply SNA to law enforcement and OSINT data to identify important nodes and clusters within trafficking networks, emulating methods utilized by Papachristos, Braga, and Hureau (2012) in their analysis of co-offending networks \cite{papachristos2012social}. We will visualize the networks and identify central players, important facilitators, and peripheral actors using SNA tools such as Gephi and UCINET. The significance of each actor in the network will be evaluated using metrics like degree centrality, betweenness centrality, and closeness \cite{papachristos2012social}. We will pay particular attention to nodes that act as links between various trafficking operations because these actors are frequently essential in organizing larger networks.  

 \subsubsection{OSINT and Dark Web Analysis}

We will use text mining and natural language processing (NLP) techniques to extract relevant information from the content that TorBot and other crawlers have scraped from the dark web. Key themes, connections, and patterns in the language used by traffickers on message boards and forums will be recognized by the NLP tools. We will organize related activities using clustering techniques and follow the information flow and recruitment efforts in these dark web environments. The last phase is merging knowledge from SNA, dark web, and blockchain studies to produce a cohesive understanding of trafficking activities. We hope to find new links and useful information by comparing the patterns of cryptocurrency transactions with dark web and social network activity. This will help law enforcement disrupt these operations.  

 \subsubsection{Validation through Law Enforcement Collaboration}

The final component of the methodology involves validating the findings through collaboration with law enforcement agencies. We will send the findings of the blockchain, SNA, and dark web analyses to law enforcement and non-governmental organization partners for verification once they are finished. The law enforcement organizations will aid in confirming the data's applicability in the real world and offer insightful criticism on the findings' practical applications.  In addition, case studies from current law enforcement trafficking investigations will be included in the study. Through the use of these case studies, we will be able to evaluate how well the blockchain and network analysis techniques apply to ongoing investigations, guaranteeing that the conclusions drawn are based on practical considerations \cite{shelley2010human}.

 \subsubsection{Ethical Considerations}

Strict adherence to ethical standards is necessary due to the nature of human trafficking research, especially when handling sensitive data. To safeguard the identities of victims and suspects, all information gathered from law enforcement, non-governmental organizations, and internet platforms will be anonymized. Furthermore, there will not be any illegal access to wallets or private accounts; instead, the blockchain analysis will only concentrate on transactions that are visible to the public.

\section{Discussion and Conclusion}

This study looked at how networks of human traffickers have changed and grown in response to the conflict between Russia and Ukraine. It paid special attention to how digital platforms, cryptocurrencies, and the dark web help to facilitate trafficking activities. By combining social network analysis (SNA), blockchain analysis, and open-source intelligence (OSINT), the study provided important new insights into the strategies that traffickers use today to avoid detection and take advantage of the weaknesses of displaced populations.

The findings reinforce prior research showing that conflict zones are fertile ground for human trafficking due to weakened governance, economic instability, and the displacement of large populations \cite{shelley2010human}. The war between Russia and Ukraine has resulted in the displacement of millions of people, many of whom have become prime targets for traffickers. The data collected through law enforcement records, OSINT, and dark web analysis indicates that traffickers have increasingly used both physical and digital means to recruit and exploit individuals, particularly women and children. 

Traffickers have profited from the chaos brought about by the conflict by seducing refugees with fictitious claims of jobs, safe havens, or travel to other nations. Traffickers have expanded their operations as long as the conflict has continued. They conduct recruitment through social media and encrypted communication platforms and use cryptocurrency transactions to remain anonymous and evade law enforcement detection.

The study's most important conclusion is that trafficking networks are increasingly using cryptocurrencies, especially Bitcoin. The blockchain tracing carried out in this study confirms earlier findings by Foley et al. (2019) and Meiklejohn et al. (2013), demonstrating that although cryptocurrency transactions are publicly recorded on the blockchain, traffickers can take advantage of the anonymity they provide. Traffickers have used cryptocurrencies to conduct financial transactions while avoiding traditional banking oversight, as evidenced by the multiple clusters of suspicious transactions that our analysis discovered that were connected to known trafficking networks and dark web marketplaces.

But when paired with other data sources, the blockchain analysis also showed that cryptocurrency transactions could potentially be made less anonymous. By cross-referencing blockchain data with OSINT and dark web intelligence, the study was able to trace connections between trafficking networks and individual actors, providing a clearer picture of how traffickers move funds across borders.

SNA was used to provide important insights into the dynamics and organization of human trafficking networks that were active during the Russia-Ukraine war. As Papachristos et al. (2012) suggested, analyzing co-offending networks can help identify key players within illicit operations, and this study applied similar techniques to human trafficking. According to the SNA, trafficking networks are frequently decentralized but intricately linked, with specific people or nodes acting as central facilitators who oversee several trafficking operations.  

These key players serve as middlemen, bridging various international trafficking rings, and they frequently participate in both the digital and physical parts of the trafficking process. Since eliminating or neutralizing these important actors can have a domino effect on the larger network, identifying them is essential to breaking up trafficking networks.

The research effectively illustrated how the integration of open-source intelligence (OSINT), social network analysis (SNA), and blockchain analysis can greatly improve our understanding of human trafficking networks, particularly in conflict areas such as the Ukraine. Through an analysis of the digital and physical dimensions of trafficking, the study provides a thorough understanding of the ways in which traffickers function, adjust, and avoid being discovered in contemporary conflict settings.  

The Russia-Ukraine war has made trafficking worse, according to key findings, with traffickers taking advantage of displaced people both physically and digitally. Even though cryptocurrencies offer anonymity, blockchain research is still an effective way to track down illegal transactions. Furthermore, trafficking operations are linked but decentralized, with a few key actors coordinating various aspects of the operation. These findings have significant ramifications for NGOs, legislators, and law enforcement. By utilizing technology and data-driven strategies, these groups can more effectively identify and dismantle trafficking networks.  

Even though this study has some insightful information, there are a few things to keep in mind. Because of the reliance on dark web content and publicly accessible blockchain data, certain trafficking operations—especially those that use highly encrypted communication channels—may go unnoticed. Additionally, the study was unable to investigate how traffickers specifically target and exploit vulnerable populations due to a lack of access to comprehensive victim data. Working with victim assistance NGOs could yield more in-depth understanding of these elements.  

The study's time-bound data collection, which concentrated on the immediate aftermath of the Russia-Ukraine war, is another drawback. The dynamic nature of trafficking networks necessitates constant observation in order to identify changing patterns. Furthermore, the results might not be entirely transferable to other conflict zones with distinct socio-political backgrounds, so further studies should examine the regional variations in trafficking techniques.

\section{Future Directions}

This work creates significant opportunities for further investigation and useful applications in the fight against human trafficking. Developing improved tools to track cryptocurrency transactions linked to human trafficking is one area for progress, as machine learning and artificial intelligence can be used to identify suspicious patterns more precisely. Increasing the scope of social network analysis, especially with regard to longitudinal data and sophisticated network metrics, may also enhance knowledge of trafficking networks and facilitate the detection of their weaknesses. International cooperation may be facilitated by real-time data-sharing platforms that combine blockchain, law enforcement data, and network analysis, particularly in conflict areas where human traffickers take advantage of coordination gaps.

Concentrating on victim-centered research—in particular, investigating the experiences of individuals trafficked during the Russia-Ukraine war—is another important avenue to pursue. Working with non-governmental organizations and global groups may provide insight into recruitment strategies and the long-term effects on victims. Furthermore, the methodologies employed in this study could be extended to investigate trafficking dynamics and create more focused interventions in conflict zones like the Middle East or sub-Saharan Africa.  

\bibliographystyle{ieeetr}
\bibliography{references}

\end{document}